# How is Software Reuse Discussed in Stack Overflow?

Eman Abdullah AlOmar[a*], Anthony Peruma[b], Mohamed Wiem Mkaouer[c], Christian Newman[c], Ali Ouni[d]

[a]*Stevens Institute of Technology, Hoboken, NJ*, [b]*University of Hawaii, Honolulu, HI*,
[c]*Rochester Institute of Technology, Rochester, NY*, [d]*ETS Montreal, Montreal, Canada*

**Abstract**

Software reuse is a crucial external quality attribute targeted by open-source and commercial projects. Despite that software reuse has experienced an increased adoption throughout the years, little is known about what aspects of code reuse developers discuss. In this paper, we present an empirical study of 1,409 posts to better understand the challenges developers face when reusing code. Our findings show that 'visual studio' is the top occurring bigrams for question posts, and there are frequent design patterns utilized by developers for the purpose of reuse. We envision our findings enabling researchers to develop guidelines to be utilized to foster software reuse.

## 1. Introduction

Modern-day software development requires the creation of products that meet quality standards and are also released in a timely manner. In order to meet these expectations, developers tend to share already implemented components or pieces of code from one project to another to avoid having to implement the same thing twice [17]. Similarly, the usage of third-party libraries, APIs, and frameworks have allowed them to reduce the time it takes to create new software. The practice of creating reusable software and utilizing already existing code to increase productivity is often referred to as Software Reuse [13]. Other than helping achieve faster development time, software reuse helps in minimizing efforts and costs. It can also aid in increasing the maintainability and portability of the system [22]. Therefore, a plethora of studies has been advocating for software reuse through good programming principles, design patterns, libraries, and frameworks. While there are several reuse principles and guidelines in theory, little is known about what prevents their wide adoption in practice. Thus, it is important to understand the main challenges developers face when it comes to reusing software, how developers perceive software reuse in practice. Utilizing already built software can also come at the cost of security, since if the reused software has flaws, those will be added to the new system being constructed, which is why, although it is recommended to reuse whenever possible, it should be done carefully.

Most quality attributes and aspects of the software development process present challenges in order for them to be achieved; reusability is not the exception. Online forums are a common way for developers to resolve unexpected coding or design problems. One of the most used sites where developers converge to ask and answer questions covering a multitude of topics related to programming, software engineering, and related areas is Stack Overflow (SO). Since its conception in 2008, SO has seen a yearly increase in its user base, with over 10 million users as of 2019. Because of the growing importance of Stack Overflow as a community of developers, many studies have been performed to understand the ways developers discuss different topics, and aspects of Stack Overflow and the community themselves.

In this study, we utilize Stack Overflow to understand the challenges that drive developers to ask questions about and discuss software reuse. To do this, we mine from Stack Overflow all questions that are tagged as *software reuse*. Our findings provide an investigation regarding the common reuse topics the developers discuss in Stack Overflow. In particular, it focuses on the identification and classification of questions and discussions related to design patterns. Finally, it provides some observations extracted from the manually analysed questions.

## 2. Study Design

The main goal of this work is to obtain and share insights with the Software Engineering community regarding how software reuse is discussed in practice by analyzing Stack Overflow (SO) posts. Figure 1 depicts an overview of our methodology. We made our dataset publicly available [21].

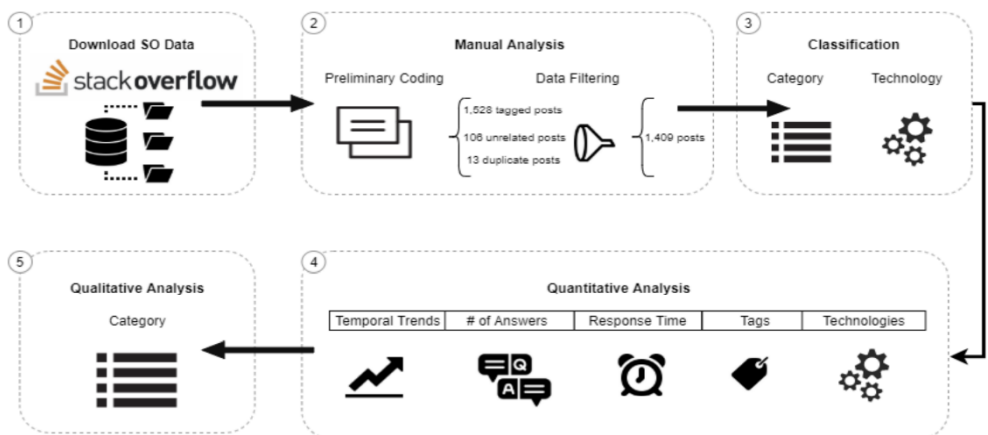

Figure 1. Overview of the study methodology.

### 2.1. Data Extraction

To conduct our study, we plan on extracting our data from Stack Overflow. The data should capture all the necessary information needed for our study, starting with questions, answers, discussions, and all their corresponding metadata (publication date, authors, TAGs, source code attached, etc.). We start the mining process through the Data Explorer web interface which releases a regularly updated data dump of the SO network's content. We considered all questions, since Stack Overflow creation, from August 2008, all the way to December 2019.

To distinguish reuse-related questions, we started with using the tags "*reuse*" and "*reusability*" as they are intuitive. We have avoided using any extra tags mainly because it has been shown that developers misuse tags in an attempt to increase views, and a large number of tags tend to cause confusion among posters [9]. Additionally, in order to mitigate tag misuse issues, the relevancy of the posts and tags were verified during the manual analysis described in Section 2.3, so that we only keep posts related to software reuse. This process consisted of manually reading the title and question description for each post to assess its relevancy. When the description is ambiguous or short, we refer to any additional comments made to the question, and then we refer to the proposed answers, which may provide a clearer description of the problem. For instance, we found multiple posts where the question was vague, but the answers were clearer based on assuming what the question poster meant, and these assumptions were validated by the poster by either choosing one as a correct answer or by adding comments to further clarify their question.

## 1.1. Manual Analysis

*Posts Filtering*. We included all posts that referred to the reuse of code and reuse of different resources that are relevant to any of the stages of the Software Development Lifecycle. In addition, we also included questions that were related to general aspects of software reusability. We excluded all questions that were not related to the aforementioned aspects. Two authors have performed the manual verification of the questions. This process consisted of manually reading the title and question description for each post to assess its relevancy. When the description is ambiguous or short, we refer to any additional comments made to the question, and then we refer to the proposed answers, which may provide a clearer description of the problem. For instance, we found multiple posts where the question was vague, but the answers were clearer based on assuming what the question poster meant, and these assumptions were validated by the poster by either choosing one as a correct answer or by adding comments to further clarify their question.

*Duplicate Posts Removal*. During the manual validation, we found one duplicate post, 10 posts that were removed (upon looking for them in Stack Overflow), and 2 posts when accessing them redirected to unrelated questions.

To summarize, our data extraction process initially mined 1,528 reuse-tagged posts, from which 106 were unrelated to reuse, and 13 were duplicates. After the filtering process, we retained 1,409 posts, created between August 2008 and December 2019.

## 1.2. Manual Classification

The classification process consists of reading *title, questions, answers* and *comments* to classify them based on various aspects, including the technology used, programming language, and category of reuse. The reuse categories were extracted from the systematic literature review, performed by Ahmaro et al. [2], on papers published from 1977 to 2013 on the different concepts surrounding Software Reusability. In this work, Ahmaro et al. [2] identified the following categories: *design patterns, component-based development, application frameworks, legacy system wrapping, service-oriented systems, application product lines, commercial off-the-shelf integration, program libraries, program generators, aspect-oriented software development and configurable vertical applications*. This set of categories was also supported by Younoussi et al. [25] whose work consisted of another systematic literature review on All about Software Reusability. During the classification process, we combined questions related to components, services, and aspects into Development, we also combined program libraries, application frameworks, and *commercial off-the-shelf integration* into *Third Party Systems*. We also noticed the existence of questions that do not fit into the existing categories, so we have added two categories, namely *Refactoring*, which can be like design patterns, except that developers discuss refactoring opportunities without explicitly mentioning any patterns. The second added category was *General*, where questions were open-ended, and their answers can belong to various categories.

Our categorization is not mutually exclusive, meaning that one post can be associated with more than one category at the same time. We also classified the posts regarding the technologies involved in the question. We investigated what programming language, framework and operating systems. Although most of the times these technologies were identified by the posters using the Stack Overflow provided tags, we still focus on the content of the question as the source for this classification due to previously identified issues involving tags usage. When looking at the content of the question, we not only referred to explicit mentions of the technologies, but also at implicit ones. Questions that we were not able to associate to specific technologies, were assigned with N/A (Not Available) but still considered for the category's classification.

## 3. Results

In this section, we present the results for each of the defined specific research questions.

*3.1. How have reusability posts grown throughout the years?*

This research question examines the growth of reusability posts, questions, and answers on Stack Overflow over the years. The purpose of this RQ is to understand the extent to which developers require help and advice on reusability-related problems and how they receive assistance they seek.

Figure 3 shows the temporal trends regarding the number of posts for each year. As can be seen, there is an increasing trend from 2008 to its peak in 2012. From there until December 2019 when the data was collected there has been a decrease in the number of posts except for 2016 which had more posts than the previous year. This trend suggests that due to the wide adoption of code reuse and how it is more and more taught as a fundamental programming practice, developers don't need much help to achieve software reuse as compared with other aspects of software engineering. It is hard to determine the reason for the peak in 2012, because as our *complete dataset* shows, almost all technologies presented an increase in their numbers for this year. However, when looking at the original tags that the posters tagged their questions with iOS showed an increase from 7 posts in 2011 to 25 in 2012. Furthermore, when querying the complete *SO* data dumps to search for the temporal trends of the usage of the "*iOS*" tag, we can see that the number of posts from 2011 to 2012 almost doubled (from 41,139 to 80,276), an increase that was not seen for any other two consecutive years. This might be in relation to new features released by Apple regarding their development kit in the early months of 2012 or during 2011 (as some technologies might take some time to be fully adopted by the developers). We also observe that questions and accepted answers share a similar pattern between the year-over-year growth for questions and accepted answers. The number of non-accepted answers slightly fluctuate throughout the year.

Figure 2 depicts a yearly breakdown of reusability questions, with and without an accepted answer. As can be seen, other than the period between 2016 and 2019, the number of questions with an accepted answer outnumber questions without an accepted answer. As the years progress, the number of questions with accepted answers decreases while questions without accepted answers increases.

*3.2. What are the tags that are associated with reusability questions?*

Figure 4 shows the top 10 popular tags associated with reusability posts. We observe that 816 questions containing the `reusability' tag, and 760 questions containing the `code-reuse' tag. The top eight tags were all related to programming languages or mobile frameworks - Java (143 post), C\# (135 posts), Android (120 posts), JavaScript (107 posts), iOS (94 posts), PHP (69 posts), C++ (65 posts), and Python (59 posts).

Next, we look at the yearly growth of the eight most popular programming languages' reusability tags in the dataset. In Figure 4, the volume by which questions associated with each tag either grows or shrinks year-by-year. From the graph, other than the year of 2011, we see that questions tagged with `reusability' and `code-reuse' share a similar pattern. We also notice that questions tagged with Java, C#, and Android show a steep increase between 2010 and 2020, while at the same time, we observe the volume of JavaScript slightly decrease, the volume of iOS has steep increase in 2012, and the volume of PHP, C++ and Python tagged posts being more-or-less constant.

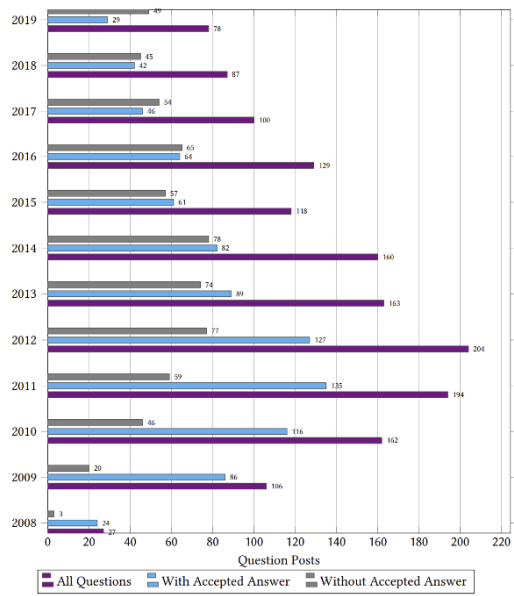

Figure 2. Count of reusability questions and answers, per year.

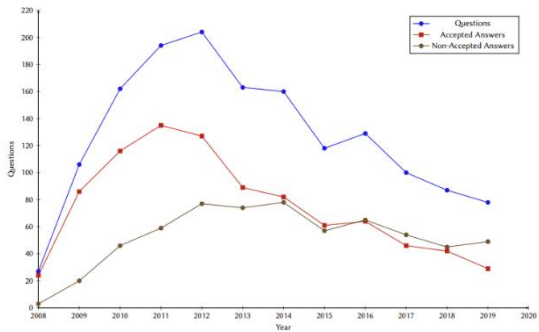

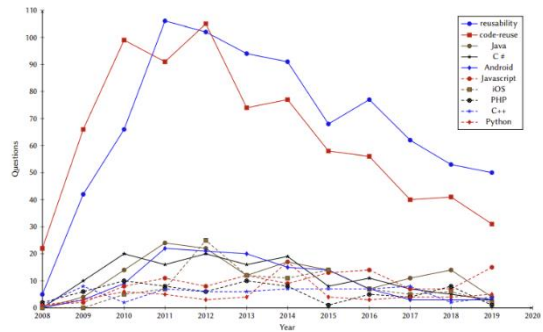

Figure 3. Temporal trend of software reuse posts.	Figure 4. Yearly growth of the popular tags associated with reusability posts.

### *3.3. What are the frequent terms utilized by developers in reusability discussions?*

To answer this question, we leverage natural language processing techniques, more specifically bigram analysis, to extract the frequent bigrams developers utilize in describing their reusability problem. Before our extraction, we first performed a series of text preprocessing activities on the post's body. These activities ensure that the text is optimized for our specific analysis [7]. The activities we perform include expansion of word contractions, removing URLs, code blocks, stop words, alphanumeric words, and punctuation, retaining only nouns, verbs, adjectives, and adverbs, and lemmatization of words. In addition to standard English stop words (e.g., `a', `and', `the', etc.), we also utilize a set of custom stop words exclusive to our reusability dataset. To derive the set of custom stop words, we generated and manually analyzed the set of frequently occurring words in our corpus. Examples of custom stop words include `thanks', `question', `answer', etc. The complete set of stop words are available in our replication package.

Table 1. Top ten frequency occurring bigrams for question posts.

| Bigram | Count | Percentage |
| --- | --- | --- |
| Visual studio | 37 | 1.13% |
| Business logic | 36 | 1.10% |
| View controller | 34 | 1.04% |
| Partial view | 32 | 0.98% |
| Web application | 32 | 0.95% |
| Copy paste | 27 | 0.83% |
| Base class | 27 | 0.83% |
| Web service | 26 | 0.79% |
| Table view | 24 | 0.73% |
| Aspnet mvc | 22 | 0.67% |
| *Others* | 2,975 | 90.95% |

As can be seen in Table 1, the occurrence of `visual studio' should not be surprising as this IDE is fast becoming a popular choice among the developer community due to its ability in supporting multiple programming languages. As part of providing context (e.g., development environment) around the reusability challenge, developers mention the IDE they utilize. Further, examination of the popular bigrams shows that developers frequently face reusability challenges related to web development as evident by the bigrams `web application', `web service', and `aspnet mvc'. Additionally, we also observe that developers face challenges around architecture/design (e.g., `business logic' and `view controller'). Finally, the bigram `copy paste' also shows how the user duplicates reusable code.

In a further analysis of the extracted bigrams, two authors manually analyzed these terms with the aim of grouping related bigrams. As part of the annotation process, each author annotated the bigrams individually and then discussed and resolved any annotations that conflicted. Our analysis yields seven groups -- Code Component (35.01%), Environment (8.93%), Framework/Library (5.04%), Architecture/Design (2.16%), Code Reuse Rationale (3.17%), Programming Language (1.01%), and Other (44.67%).

Categorized under Code Component, we observe terms that are related to parts of the source code, such as `base class', `switch\_statement', and `static\_method'. Under Environment, we see references to tools/IDE's and directories/file types like `sql\_server' and `xml\_file', respectively. Bigrams categorized as Framework/Library include those that are web-related (e.g., `ruby rail' and `javascript jquery') and other specialized libraries (e.g., `entity\_framework', `twitter api', and `facebook api'). Under Architecture/Design, we see references to patterns such as `dependency\_injection', `factory\_pattern', etc. The bigrams under Code Reuse Rationale relate to actions or situations that lead developers to seek reusability assistance; these include `hard\_maintain' ,`multiple\_place', `avoid repeating', etc. The bigrams under Programming Language refer to instances where the developer mentions the language for which they require assistance. The low volume of such bigrams is due to developers preferring to utilize the tag feature to indicate the programming language. Bigrams that did not fall into any of the other six categories are listed as Other; such terms include `http request', `add reference', `open source', etc.

*3.4. What are the frequent design patterns utilized by developers for the purpose of software reuse?*

Our first investigation aims to understand which Design Patterns (DP) are popular among developers and highly discussed in SO. So, we start with reporting the distribution of questions with respect to the corresponding DP being asked. Then we compare all DPs belonging to the same category. Figure 5a is associated with creational design patterns. As we can see, the Singleton pattern with being the main 47.8% design pattern that almost encapsulates half of the questions. The Builder design pattern comes second with only 16.4%, but compared to the other design patterns including Prototype, Abstract Factory, Dependency Injection, Factory Method, it takes the second largest category. Questions targeting Structural DPs seems to be more evenly distributed. According to Figure 5b, only the Decorator DP has a significantly large number of questions compared to the remaining DPs, e.g., Adapter, Composite, Facade, Proxy, whose categories range between 6.9% and 16.7%.    For behavioral DPs, we noticed that there are two categories, the first one contains five design patterns, namely the State pattern, the Strategy, Observer, Command, and Visitor. These DPs have taken most of the questions with over 80%. The remaining behavioral DPs, including Mediator, Iterator, Memento, and Interpreter, seem to be not popular since all of them contribute with less than 5% of questions as seen in Figure 5c.

When contrasting the number of questions per DP with the average time needed to get an accepted answer, we have noticed that popular patterns, with a high number of questions, are not necessarily the "hardest" to respond to. When looking at Figures 6a, 6b, 6c, prototype and Dependency injection take a significant time before reaching an accepted answer. Similarly, while Decorator takes the lead in the number of questions, it is at the bottom, especially when compared with Bridge or Proxy.

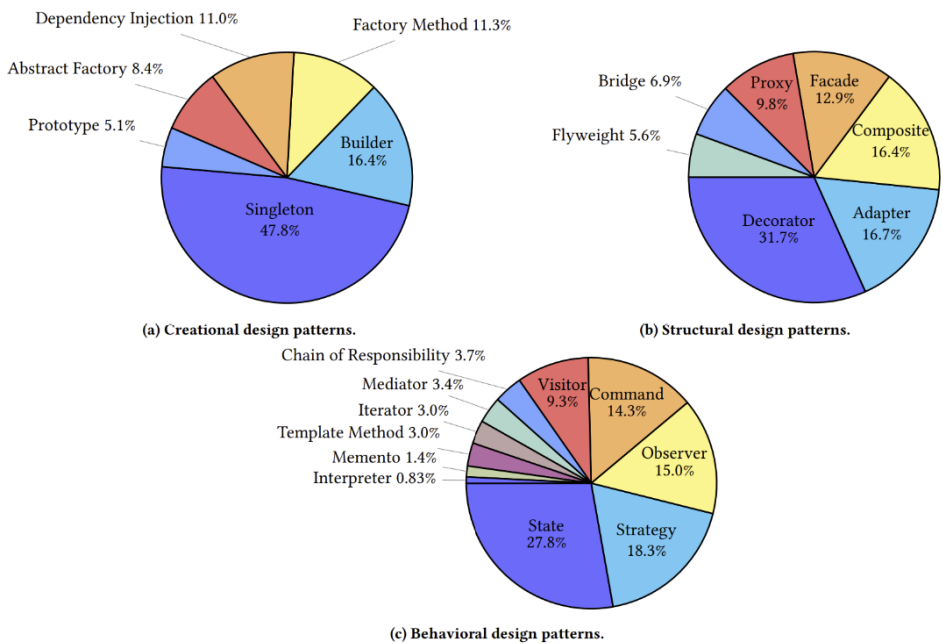

Figure 8. Breakdown of the volume of instances for each design pattern category.

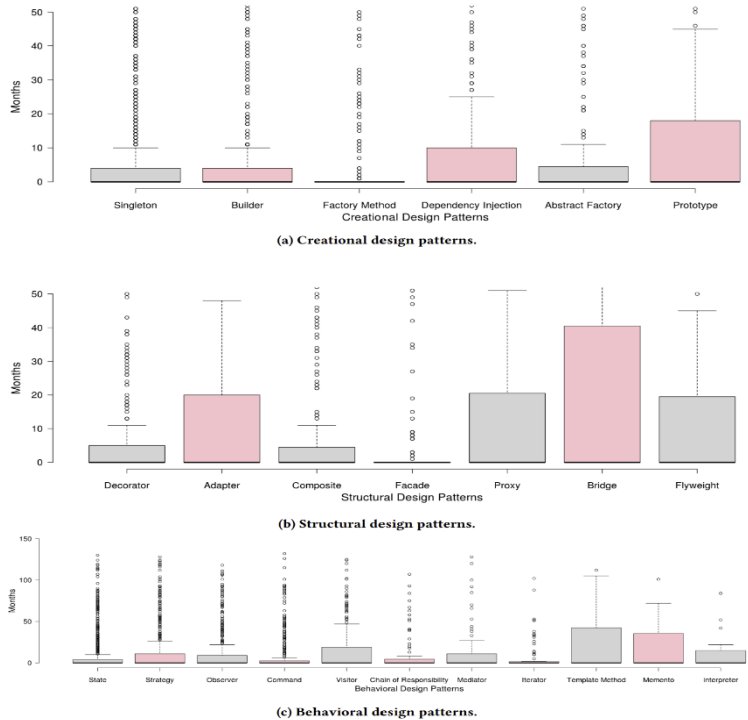

Figure 6. The average time needed to get an accepted answer.

## 4. Observation

While performing our manual classification, we also noted that comments and observations regarding the analyzed posts. We present, with extracts from the original posts, some of these observations. For each example, we also include the QuestionIDs (QID)

**Observation**. As we expected, several questions were asked because developers wanted to avoid creating the needed code themselves. Although these underlying needs were also about achieving code reuse, this shows that developers also look into ready-to-go algorithms from the community that can be reused for their own needs (QID: 1234514):

``Is there a very simple map-base code snippet in C out there that I could use to do the caching? (I just don't want to reinvent the wheel). [...]''

**Observation**. Reusability is not only about reusing large blocks of code, we noted a number of questions that discuss reusability in terms of reusing variables simply to avoid the creation of new ones. For example (QID: 3005259)

``[...] Here, I have a *writer* which archives the object *me*. Then, I use another *CArchive* object reader to un-archive it. Is it possible to re-construct or set any property of *writer* to make it, the reader instead of declaring another *CArchive* object *reader*?''

**Observation**. Reusability is discussed targeting company level practices. Questions regarding how to achieve an organization wide reusability culture and how companies go about reusability are asked. This shows an interest from the industry workers to increase productivity in their companies by performing reusability (QID: 1475316):

``Frequently when we introduce a new feature into an application we may produce artifacts, such as useful methods or classes that could be reused in other areas of our applications. [...] Since we often develop in teams, it is important to share these pieces of code to prevent rework and duplication. [...] I am interested in best practices and proven techniques around documentation, code diagrams, meetings(?) to ensure code is reused correctly. [...]''

**Observation**. Developers typically discuss best practices to share methods, across various, but related projects. They typically structure them in the form of libraries, and offered via APIs. For instance, this question concerns the creation of shared internal libraries (QID: 1773634):

``I've noticed in pretty much every company I've worked that they have a common library that is generally shared across a number of projects. [...] What are the things you have in your common library and more importantly how do you structure the common libraries to make them easy to improve and incorporate across other projects?''

**Observation**. A good architecture for any given software product could be said to be the one that better handles the trade-offs between the non-functional requirements. We noted some questions were asked regarding the priority of different quality attributes. In this example, QID: 523222, they discuss maintainability, performance and reusability:

``[...] All that being said, I would order: Maintainability (Change is a fact of life.), Performance (No one likes to stare at the hourglass.), Reusability (Difficult to determine what code should be used again.).''

**Observation**. One aspect that one would think developers should be more concerned about when trying to reuse code or libraries found on the web, is the ethics of it and whether it is legal or not. From our dataset, only one question discussed this (QID: 11023772):

``Say I saw tutorial online on doing something with JavaScript or jQuery on a web page. Is using the code on my commercial site bad manners or even illegal? I slightly modified it but the important stuff is from the tutorial.''

**Observation**. Reusability is discussed not only in terms of reusing code, but also reusing different elements around the whole process of developing software. For example, developers ask regarding the reuse of images. As images tend to be of large sizes, they seek options to avoid having to upload the same image multiple times to their server as a way of decreasing memory load and increasing performance. For example, QID: 9334780 raised the below concern:

``[...] There has to be some way to reuse an image without reuploading it for every single post I need to use it on. I'm nauseous at the thought of 54 instances of the same image being present on my server. [...]''

## 5. Related Work

As Stack Overflow sees growth in its user base and importance in the Software Engineering and related areas community, researchers started to perform studies gathering information from this site. Some of these works study the utility of Stack Overflow as a Q&A site to offer support to developers and discuss the shift of developers from other discussion forums to it [10,23]. Other studies have focused on the functionalities provided by Stack Overflow and general aspects of the site to provide more insights regarding the community. Previous studies have focused on analyzing specific topics using Stack Overflow for several domains such as machine learning, code smell, security, mobile application, and code reuse [5]. Software reusability has been a topic of interest since 1970s, and because of that, a large number of literatures that discussed it is available. This section, although not accounting for all available work, does its best in presenting multiple papers from the different ways reusability is discussed. Previous work discussed general aspects of reusability, identifying challenges, topics, issues and principles [2,4,8,12,15,18,24,25].

AlOmar et al. [4] studied how developers refactor their code to improve its reuse by analyzing the impact of reusability refactorings on the state-of-the-art reusability metrics. Ahmaro et al. [2] conducted a systematic literature review to identify the definition, approaches, benefits, reusability levels, factors, and adaption of software reusability. The authors found that the concept of software reusability comprised of 11 approaches, namely, design patterns, component-based development, application frameworks, legacy system wrapping, service-oriented systems, application product lines, COTS integration, program libraries, program generators, aspect-oriented software development and configurable vertical applications. A study is reported on the relationship of complexity and reuse design principles is reported by Anguswamy and Frakes [8] Their findings show that the higher the complexity the lower the ease of reuse. Lubars et al. [15] contrasted code reusability in the large versus code reusability in small with regards to several aspects, including, size, complexity, application, and problems associated with locating and reusing the code. The author highlighted that code reusability in the small has had limited impact because of its strongly self-centered orientation, whereas code reusability in the large has had limited impact because of its high degree of difficulty in finding the reusable components. In a similar context, Mockus [18] performed large-scale code reuse study in open-source software and found that more than 50% of the files were used in more than one project. Yin and Lee [24] conducted a survey to examine the characteristics of software reusability from the points of view of software engineering as well as knowledge engineering. Younoussi and Roudies [25] presented a systematic literature review on software reusability. They pointed out that few studies examined barriers of reusability, and organizations need to adapt software reusability approaches. Reusability and code reuse are also discussed in relation to specific topics [1,6,11,14,16,19,20, 22]. Lotter et al. [14] explored code reuse between Stack Overflow and Java open-source systems in order to understand how the practice of reusing code could affect future software maintenance and the correct use of license. Their findings show that there is up to 3.3 % code reuse within Stack Overflow, while 1.0 % of Stack Overflow code is reused in Java projects. Patrick [19] investigated reusability metrics with Q&A forum data. The author proposed an approach (LANLAN), using word embeddings and machine learning, to classify Q&A forum posts into support requests and problem reports, as well as reveal information in relation to software reusability and explore potential reusability metrics. In another context, Abdalkareem et al. [1] performed an exploratory study on 22 Android apps to explore how much, why, when, and who reuses code. They found that 1.3 % of the Apps were constructed from Stack Overflow posts, and discovered that mid-aged and older apps reuse Stack Overflow code later in their lifetime. An et al. [6] also explored Android apps and found that 15.5 % of the apps contained exact code clones, and 60 out of 62 apps, had potential license violations.

## 6. Conclusion

In this paper, we presented insights regarding how developers discuss software reuse by analyzing Stack Overflow. These findings can be used to guide future research and to assess the relevancy of software reuse nowadays. Our findings show that software reuse is a decreasing trend in Stack Overflow which might indicate that developers have widely adopted this practice and thus few questions regarding it emerge as it is well grasped by the community. On the other hand, they might indicate that, since software reuse is so deeply embedded in achieving good quality software and is often taught as an essential quality attribute, it has spread to a wide variety of topics and thus the questions are not posted solely as reusability or code reuse issues. Additionally, our findings show that 'visual studio' is the top occurring bigrams for question posts, and there are frequent design patterns utilized by developers for the purpose of reuse. Our study opens the door for software reuse researchers to further understand the software reuse challenges. In the future, we plan to examine developers' discussion from other forums to draw more generalizable conclusions. We also plan to extend our work to consider reuse in specific contexts.